# COMBINATIONAL PROCESSES IN DENSE HIGH-TEMPERATURE PLASMAS

Yu. V.Arkhipov, F.B.Baimbetov, A.E.Davletov,

O.S.Puchinkina, O.V.Salov, K.V.Starikov, V.V.Voronkov

Let a fully ionized plasma consist of electrons and ions. Concentration of these particles in the plasma varies in the range n=$n_e$+$n_i$=$10^{20}$-$10^{25}$ sm$^{-3}$, and the temperature $10^5$-$10^7$ K. Hereinafter all numerical calculations concern to high-temperature, hydrogen, dense plasma, that is typical for the main sequence stars. Pressure of such a plasma is nearly $10^{11}$Bar. Under these conditions te nonideality and density parameters are close to one, for example, for the sun $r_s$~0.4 and Γ~0.1. Consequently, polarization and quantum-mechanical effects of diffraction and symmetry contribute significantly to such plasma properties.

In the context of dielectric response theory one can find the interionic psevdopotential, which accounts for the local field correction and shielding effect [2]:

$$\widetilde{\Phi}_{ii}(k) = \widetilde{\varphi}_{ii}(k)\left(1 - \frac{\widetilde{\varphi}_{ee}(k)}{\frac{k_B T}{N_e} + \widetilde{\varphi}_{ee}(k)(1 - G_{ee}(k))}\right), \qquad (1)$$

where $G_{ee}$ is the electron-electron local field correction function, $\widetilde{\varphi}_{ee}(k), \widetilde{\varphi}_{ii}(k)$ represent the Fourier transforms of the Deutsch micropotential [1].

In contrast to the Coulomb and Debye – Hukel potentials pseudopotential (1) is restricted at the origin like the Deutsch potential. It is explained by the fact that this pseudopotential takes into account quantum-mechanical effects.

At short interparticle distances and small values of the nonidelity parameter Γ pseudopotential (1) goes close to the Deutsch potential and approaches the Debye – Huckel potential at large interparticle distances as both take into account electron shielding of ions.Such a situation changes at large Γ values because of the local field correction..

With the help of the linear dielectric response theory one can find the structure factors $S_{ee}(k), S_{ei}(k), S_{ii}(k)$, which take into consideration the local field correction and quantum-mechanical effects as well [2,3]. Knowledge of the structure factors allows one to get the Coulomb logarithm of the form [4]:

$$L_C = \int_0^\infty \frac{dk}{k} \frac{\left[S_{ee}(k) S_{ii}(k) - S_{ei}^2(k)\right]}{(1 + k^2 \lambda_{ei}^2)^2}. \qquad (2)$$

In the one component plasma the electron equation of motion with the electron-ion collisional friction, reads as

$$m\ddot{\vec{r}} = -\sigma \vec{r} - m v_e \dot{\vec{r}} - e \vec{E}, \qquad (3)$$

where $v_e$ refers to the effective collision frequency that is to be calculated as follows [5]:

$$v_e = \frac{4}{3\sqrt{2\pi}} \frac{e^4 n}{\sqrt{m}(k_B T)^{3/2}} L_C. \qquad (4)$$

The first term in equation (3) is quasi-elastic force caused by the charge fluctuations, the second term is due to the friction; the last one is the external electromagnetic field.

Solving equation (3) and taking into consideration the relation $\sigma = \frac{e^2 n}{\varepsilon}$, one arrives at:

$$\vec{r} = -\frac{e\vec{E}}{m}\left[\frac{w_{Le}^2 - w^2}{(w_{Le}^2 - w^2)^2 + v_e^2 w^2} - i\frac{v_e w}{(w_{Le}^2 - w^2)^2 + v_e^2 w^2}\right], \quad (5)$$

where $\vec{E} = \vec{E}_0 e^{iwt}$, $w_{Le} = \frac{e^2 n}{\varepsilon_0 m}$ denotes the plasma frequency.

Taking into consideration that the vector $\vec{E}$ is parallel to the vectro $\vec{r}$ one obtains the following formula for the inductive capacity $\alpha$

$$\alpha = -\frac{ner}{\varepsilon_0 E},$$

that is a complex quantity and, thus, is represented as $\alpha = \alpha' + i\alpha''$

$$\alpha' = \frac{w_{Le}^2(w_{Le}^2 - w^2)}{(w_{Le}^2 - w^2)^2 + v_e^2 w_{Le}^2}, \quad (6)$$

$$\alpha'' = -\frac{v_e w}{(w_{Le}^2 - w^2)^2 + v_e^2 w_e^2}. \quad (7)$$

After substitution $u = w/w_{Le}$, $q = v_e/w_{Le}$ Eqs.(6) and (7) are rewritten as

$$\alpha' = \frac{1-u}{(1-u)^2 + q^2 u^2}, \quad (8)$$

$$\alpha'' = -\frac{qu}{(1-u)^2 + q^2 u^2}. \quad (9)$$

Here $q = \Gamma^{3/2} L_C$ (see Eq.(5)).

The real and imaginary parts of the refractive index depend on the dielectric capacity as:

$$\eta' = \frac{\alpha''}{\sqrt{2(\sqrt{(1+\alpha')^2 + \alpha''^2} - 1 - \alpha')}}, \quad (10)$$

$$\eta'' = \sqrt{\frac{\sqrt{(1+\alpha')^2 + \alpha''^2} - 1 - \alpha'}{2}} \ . \tag{11}$$

Numerical results are presented below for different values of the nonideality parameter $\Gamma$.

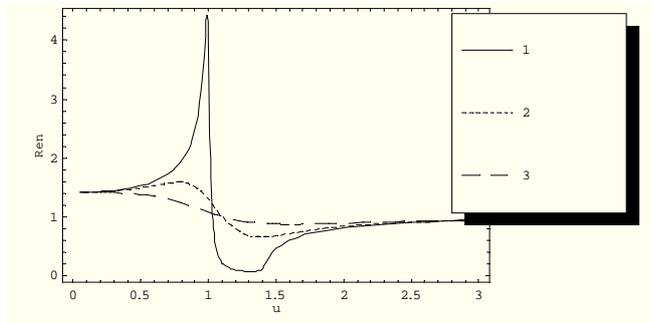

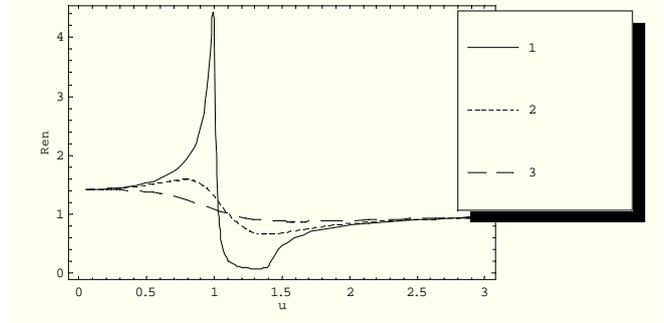

Fig. 1

Fig. 2

The real part of the refraction index against the parameter $u$ in the random phase approximation.

1 — $\Gamma=0.1$
2 — $\Gamma=0.5$
3 — $\Gamma=0.9$

The imaginary part of the refraction index against the parameter $u$ in the random phase approximation.

1 — $\Gamma=0.1$
2 — $\Gamma=0.5$
3 — $\Gamma=0.9$

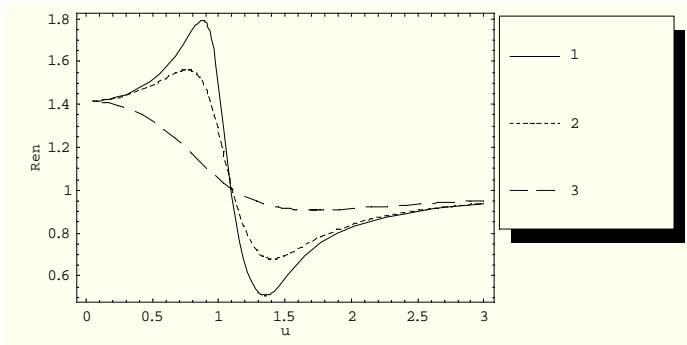

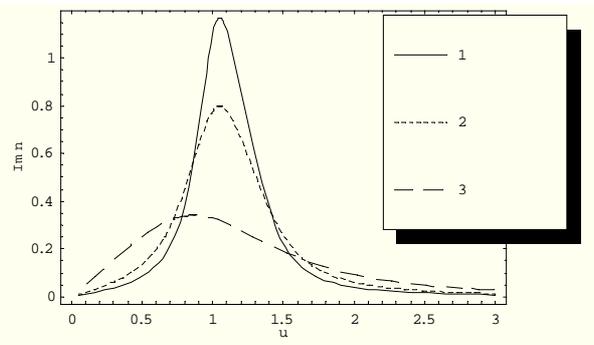

Fig. 3

Fig. 4

The real part of the refraction index against the parameter $u$ in the local field correction approximation.

1 — $\Gamma=0.1$
2 — $\Gamma=0.5$
3 — $\Gamma=0.9$

The imaginary part of the refraction index against the parameter $u$ in the local field correction approximation.

1 — $\Gamma=0.1$
2 — $\Gamma=0.5$
3 — $\Gamma=0.9$

On the basis of the figures one can canclude the following:

1. Range of frequencies with abnormal dispersion grows when the nonideality parameter increases.
2. The maximum of the absorption curve shifts to lower frequencies.
3. The height of absorption curve peak decreases when the nonideality parameter grows, and spectral width enlarges.

Therefore the quantum-mechanical and nonideality effects play significant role at high plasma densities.